# CENI: a Hybrid Framework for Efficiently Inferring Information Networks

Qingbo Hu[*], Sihong Xie[*], Shuyang Lin[*], Senzhang Wang[†], Philip S. Yu[*]

[*]University of Illinois at Chicago, [†]Beihang University

{qhu5, sxie6, slin38, psyu}@uic.edu, szwang@cse.buaa.edu.cn

## Abstract

Nowadays, the message diffusion links among users or websites drive the development of countless innovative applications. However, in reality, it is easier for us to observe the timestamps when different nodes in the network react on a message, while the connections empowering the diffusion of the message remain hidden. This motivates recent extensive studies on the ***network inference problem***: unveiling the edges from the records of messages disseminated through them. Existing solutions are computationally expensive, which motivates us to develop an efficient two-step general framework, ***Clustering Embedded Network Inference*** (CENI). CENI integrates clustering strategies to improve the efficiency of network inference. By clustering nodes directly on the timelines of messages, we propose two naive implementations of CENI: ***Infection-centric CENI*** and ***Cascade-centric CENI***. Additionally, we point out the ***critical dimension*** problem of CENI: instead of one-dimensional timelines, we need to first project the nodes to an Euclidean space of certain dimension before clustering. A CENI adopting clustering method on the projected space can better preserve the structure hidden in the cascades, and generate more accurately inferred links. This insight sheds light on other related work attempting to discover or utilize the latent cluster structure in the disseminated messages. By addressing the critical dimension problem, we propose the third implementation of the CENI framework: ***Projection-based CENI***. Through extensive experiments on two real datasets, we show that the three CENI models only need around 20% ∼ 50% of the running time of state-of-the-art methods. Moreover, the inferred edges of Projection-based CENI preserves or even outperforms the effectiveness of state-of-the-art methods.

## Introduction

Nowadays, with the assistance of online social media and social networks, a piece of information has the potential to be widely spread within a short period of time. In real-life, industrial companies utilize such fact to develop innovative applications, such as marketing campaigns for new products through online social networks and monitoring trendy topics to profile potential customers. During such process, many interesting problems emerge, which attracts both computer and social science researchers.

## Network Inference Problem

One of the most fundamental problems related to online social networks is to infer the information transmission capacity between two users by analyzing their online activities. The solution to this problem makes it possible for us to develop efficient strategies to promote a product through online social networks, a.k.a. *viral marketing* (Kempe, Kleinberg, and Tardos 2003). In general, we name a message that diffuses through networks as a ***contagion*** or ***cascade*** and those users who react on the message as ***infected*** users. In the ideal case, cascades are able to reveal the links among network users, if we can fully observe the information flow. In other words, if every user explicitly states from whom s/he receives the message when s/he reacts on it, we are aware of the connection between him/her and his/her information source. However, this is far from the truth in reality.

In real life, the information flow is usually partially observable or completely unobservable. For example, the connections in an online social network might not be entirely visible due to the privacy settings. In a more challenging scenario, a blog user may never mention his/her information source when s/he posts a blog. Fortunately, during the propagation of a cascade, it is usually easy to know the time when each infected user receives the contagion. As a result, how to uncover the information connections among users from their observed infection time to different cascades, a.k.a. ***network inference problem***, becomes an important topic and has received extensive attention from researchers (Gomez-Rodriguez, Leskovec, and Krause 2010; Gomez-Rodriguez, Balduzzi, and Schölkopf 2011; Gomez-Rodriguez, Leskovec, and Schölkopf 2013; Du et al. 2013; Myers and Leskovec 2010; Wang, Ermon, and Hopcroft 2012; Abrahao et al. 2013; Netrapalli and Sanghavi 2012; Snowsill et al. 2011; Wang et al. 2014).

## Solutions based on Survival Analysis

Recently, methods based on survival analysis to solve the network inference problem have burgeoned (Gomez-Rodriguez, Balduzzi, and Schölkopf 2011; Gomez-Rodriguez, Leskovec, and Schölkopf 2013; Du et al. 2013; Wang, Ermon, and Hopcroft 2012; Wang et al. 2014). The fundamental idea of this type of solutions is that regarding to a specific cascade and its infected users, the probability that a user is infected by a previous one monotonically de-creases with the increase of the difference of their infection time. In other words, a user is more likely to be infected by another who has a closer infection time with respect to his/her. Survival analysis based models assume this probability follows a distribution that is governed by a parameter associated with each pair of users, which can be estimated through Maximum Likelihood Estimation (MLE). The value of the parameter indicates the strength of the directed connection between two users. Since information can be passed between each pair of users in either direction, we need to estimate $N^2$ parameters, where $N$ is the number of total users in the network. As one may imagine, the large number of parameters causes efficiency problem for survival analysis based methods. For example, if two users stay far away from each other in the infection timelines of most cascades, survival analysis based solutions might still end up with assigning a weak link between them. Inferring such links brings us extra computational burdens and is usually unnecessary, since they are normally viewed as noise and being discarded later.

**Proposed Hybrid Framework**

In order to improve the efficiency of survival analysis based approaches, we may utilize the information extracted from the nodes' cluster structures. If we are aware of the clusters in the network, we might omit the connections crossing clusters when we infer the network from cascades. Similar to a divide-and-conquer technique, this may reduce the size of our problem: clusters produce the division of cascades, since we only need to consider the potential influencers within a cluster for each node. Although this idea might lose some true inter-cluster edges, we may be able to control such lost to an acceptable level with a carefully designed clustering strategy. Besides the benefit of improving the efficiency of inference, the precision of inferred edges may increase as well, since the clusters might cut off incorrectly inferred edges of traditional approaches.

However, one should notice that in the network inference problem, even the edges among nodes are not observed, not to mention the cluster structures in the network. In fact, this causes existing community detection models based on both cascades and user connections, such as (Barbieri, Bonchi, and Manco 2013), not applicable. Besides this, the purpose of adopting clustering method in the proposed framework is to discover the group structures to provide the best efficiency-effective trade-off for network inference. This is very different from the "community detection motivated" clustering models, which aim to find communities that have real-life meaning in the dataset.

To extract the clustering structure from the cascades, we adopt a simple intuition: if the difference between two users' infection time is relatively small in observed cascades, the two users might be close and should be considered in the same cluster. This indicates that the network inference method may need to spend time to estimate their connection, since the same cluster members are more likely to have strong connections. Such idea leads to the proposed hybrid framework: ***Clustering Embedded Network Inference*** (CENI). CENI is a two-step framework, which first attempts to obtain the cluster structures and then incorporate them with a network inference model to infer edges.

**Critical Dimension**

Obtaining clusters directly from the timelines of infection is usually straightforward. For instance, we can assign a window for each infected node to identify its possible cluster members. Therefore, we first propose two implementations of the CENI framework, namely ***Infection-centric CENI*** (I-CENI) and ***Cascade-centric CENI*** (C-CENI), which attempt to naively find clusters based on the timelines.

In practice, we find that although I-CENI and C-CENI only need around 20% ∼ 50% of the running time of some state-of-the-art algorithms, they are not stable for all datasets: in some cases, compared to the state-of-the-art algorithms, I-CENI and C-CENI may lose much accuracy of inferred edges. We propose a possible explanation to this phenomenon: the cluster structure hidden in the diffusion time gaps are too complicated to be captured by the one-dimensional approaches. Therefore, we may want to look for better clustering strategies on higher dimensional space. However, how to find a particular dimension of node representation so that the obtained clusters can generate high-quality inferred edges? We refer this specific dimension to the ***critical dimension*** of CENI framework. Unfortunately, it is hard to find the critical dimension without actually running the network inference algorithm, which is much more time-consuming than the clustering.

In order to conquer this challenge, we propose another CENI model: ***Projection-based CENI*** (P-CENI). P-CENI first adopts a heuristic method by using a hinge loss estimator to identify the critical dimension. Afterwards, P-CENI obtain the embedded node representations in a space of the critical dimension and then apply a clustering algorithm to find the clusters for network inference. Through substantial experiments on two real datasets, we show that P-CENI can run in a similar time as I-CENI and C-CENI, while still preserves or even improves the effectiveness of state-of-the-art network inference algorithms in terms of the F-measure of inferred edges. We list our major contributions as follows:

- To the best of our knowledge, this article is the first one to explore how to leverage latent cluster structures to improve the efficiency of network inference models.

- We propose a two-step general framework, ***Clustering Embedded Network Inference*** (CENI), to efficiently incorporate the information from derived clusters to infer information networks. We also introduce three distinct implementations of CENI: I-CENI, C-CENI and P-CENI.

- We point out the ***critical dimension*** problem of CENI framework. Moreover, we recognize a simple estimator to efficiently find this critical dimension. Such findings provide important insights for other future work attempting to discover or utilize cluster structures on cascades.

- Through extensive experiments on two real datasets, we show that I-CENI, C-CENI and P-CENI are very efficient: they only need around 20% ∼ 50% of the running time of state-of-the-art approaches. Moreover, on both datasets, the F-measure of inferred edges by P-CENI is very similar to or even better than the F-measure of inferred edges by the state-of-the-art methods.

## Clustering Embedded Network Inference

In the following three subsections. we first introduce some preliminaries and the two-phase general framework of Clustering Embedded Network Inference (CENI). The second subsection presents the implementation of the second phase, which is shared by three CENI models. At last, the third subsection introduces three implementations of the first phase of CENI, which naturally distinguish different CENI models.

### Preliminaries & CENI Framework

First of all, we introduce some important concepts and notations that are closely related to this article. We denote a network by $\mathcal{G} = (\mathcal{V}, \mathcal{E})$, where $\mathcal{V}$ is the set of nodes and $\mathcal{E}$ is the set of edges. The input of the network inference problem is the records of cascades that spread through $\mathcal{G}$. A cascade, say $c$, can be represented by a vector of length $N$: $\boldsymbol{t}^c = (t_1^c, t_2^c, ... t_N^c)$, where $t_i^c$ is the $i^{th}$ user's infection time to cascade $c$, and $N$ is the number of nodes in the network, a.k.a. $|\mathcal{V}| = N$. Additionally, we have $t_i^c \in [0, T^c] \cup \{\infty\}$, where $T^c$ is the length of the observation time window for cascade $c$, and $\infty$ simply denotes the user is not infected within the time window. Each cascade has a separate clock, and it is set to 0 at the time point when the first node is infected. Moreover, for simplicity, we usually fix the time windows of different cascades to the same length, which means $T^c = T, \forall c$. As a result, the network inference problem can be stated as how to uncover hidden $\mathcal{E}$ from known $\mathcal{V}$ and $\boldsymbol{t}^c, c = \{1, 2, ..., M\}$, where $M$ is the total number of cascades. For each node in the network, we assume there is an associated subset of nodes, which contains the members in the same cluster as itself. We denote these subsets by $\mathcal{C} = \{C_1, C_2, ..., C_N\}$, where $C_i$ is the subset of cluster members associated with the $i^{th}$ node. Moreover, in some implementations of CENI, $\mathcal{C}$ might change with respect to cascades. In such cases, we replace $\mathcal{C}$ with $\mathcal{C}^c = \{C_1^c, C_2^c, ..., C_N^c\}$ for the ease and rigorousness of statement, where $C_i^c$ denotes the cluster members of the $i^{th}$ node in cascade $c$.

As we stated previously, CENI is a two-step framework. Fig. 1 displays a sketch of the structure of CENI: the first phase derives $\mathcal{C}$ from observed cascades based on network nodes. The second phase integrates the obtained $\mathcal{C}$ to a traditional network inference model to infer $\mathcal{E}$, which also requires the observed cascades as input. The general description of CENI provides users the flexibility to try out different implementations of the clustering and inference phases. In this article, we focus on the problem of finding suitable clustering strategies to improve the efficiency of traditional network inference models while preserving their effectiveness. Therefore, we emphasize the discussion of the implementation of the first phase, which essentially leads to three distinct CENI models. Meanwhile, we fix the approach in the inference phase as the same one for different CENI models to illustrate a better clustering strategy.

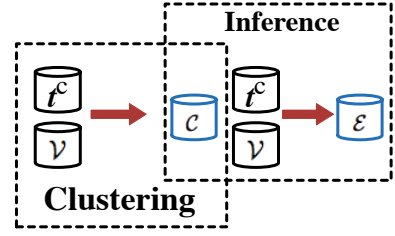

Figure 1: CENI Framework. Black cylinders denote the known information, and blue ones denote the unknown information. Dashed squares marked the two phases of CENI.

More particularly, we slightly modify the NetRate (Gomez-Rodriguez, Balduzzi, and Schölkopf 2011) model to fit in the second phase, because this model is widely used in many cases and serves as the foundation of several other existing methods (Gomez-Rodriguez, Leskovec, and Schölkopf 2013; Wang, Ermon, and Hopcroft 2012; Du et al. 2013; Wang et al. 2014; Kurashima et al. 2014).

### Implementation of the Inference Phase

If we have the cluster members for each node, we can integrate them into the *transmission likelihood* of NetRate model. The transmission likelihood, denoted by $f(t_i^c | t_j^c; \alpha_{j,i})$, refers to the conditional probability that an infected node $i$ of cascade $c$, is infected by another node $j$, given the fact $j$ is also a infected node of $c$. $f(t_i^c | t_j^c; \alpha_{j,i})$ is a function of the infection time of node $i$ and $j$ to cascade $c$, and $\alpha_{j,i}$ is a non-negative model parameter associated with edge $j \to i$. If the directed edge $j \to i$ exists, $\alpha_{j,i}$ has a positive value, otherwise $\alpha_{j,i} = 0$.

Practically, in the inference phase of CENI, we employ two types of transmission likelihood functions, i.e. an exponential distribution:

$$f(t_i^c | t_j^c; \alpha_{j,i}) = \begin{cases} \alpha_{j,i} \cdot e^{-\alpha_{j,i} \Delta_{j,i}^c} & t_j^c < t_i^c, j \in C_i \text{ or } C_i^c \\ 0 & \text{otherwise} \end{cases} \quad (1)$$

and a Rayleigh distribution:

$$f(t_i^c | t_j^c; \alpha_{j,i}) = \begin{cases} \alpha_{j,i} \Delta_{j,i}^c \cdot e^{-\frac{1}{2} \alpha_{j,i} (\Delta_{j,i}^c)^2} & t_j^c < t_i^c, j \in C_i \text{ or } C_i^c \\ 0 & \text{otherwise} \end{cases} \quad (2)$$

where $\Delta_{j,i}^c = t_i^c - t_j^c$. The above two transmission likelihood functions define that influence can only propagate forward time. In other words, an infected node will not have further influence on any node who is already infected. This is the same as the likelihood functions of the original NetRate (Gomez-Rodriguez, Balduzzi, and Schölkopf 2011). However, Eq. (1) and Eq. (2) have an additional assumption related to the cluster members: a node can only infect its cluster members. In other words, for an observed cascade, we assume the message diffuses in different clusters independently, while the inter-cluster connections are neglected for the sake of improving the efficiency.

**Survival function**, denoted by $S(t_i | t_j; \alpha_{j,i})$, refers to the probability that a node $j$ does **not** infect $i$ by time $t_i$, given that the node $j$ gets infected at time $t_j$. Thus, we have $S(t_i|t_j; \alpha_{j,i}) = 1 - \int_{t_j}^{t_i} f(t|t_j; \alpha_{j,i})dt$. **Hazard function**, denoted by $H(t_i|t_j; \alpha_{j,i})$, on the other hand, refers to the instantaneous infection rate, which is: $H(t_i|t_j; \alpha_{j,i}) = \frac{f(t_i|t_j;\alpha_{j,i})}{S(t_i|t_j;\alpha_{j,i})}$. As calculated in (Gomez-Rodriguez, Balduzzi, and Schölkopf 2011), we can show that the survival and hazard functions of a transmission likelihood taking the exponential distribution are:

$$S(t_i^c|t_j^c; \alpha_{j,i}) = e^{-\alpha_{j,i} \Delta_{j,i}^c}$$
$$H(t_i^c|t_j^c; \alpha_{j,i}) = \alpha_{j,i}$$

Similarly, the survival and hazard function of a transmission likelihood taking the Rayleigh distribution are:

$$S(t_i^c|t_j^c; \alpha_{j,i}) = e^{-\alpha_{j,i} \cdot \frac{(\Delta_{j,i}^c)^2}{2}}$$
$$H(t_i^c|t_j^c; \alpha_{j,i}) = \alpha_{j,i} \cdot \Delta_{j,i}^c$$

With the notations of survival and hazard functions, the log-likelihood of a cascade $c$, denoted by $\ell_c(\boldsymbol{t}^c; \boldsymbol{A}, \mathcal{C})$ can be readily computed[1], where $\boldsymbol{A}$ is the matrix notation of model parameters ($\boldsymbol{A}_{i,j} = \alpha_{i,j}$):

$$\ell_c(\boldsymbol{t}^c; \boldsymbol{A}, \mathcal{C}) = \sum_{\{t_i \leq T\}} \sum_{\{m:t_m^c > T\}} \log S(T|t_i^c; \alpha_{i,m}) + \sum_{\{j:t_j^c < t_i^c, j \in C_i \text{ or } C_i^c\}} \log S(t_i^c|t_j^c; \alpha_{j,i}) + \log \left[ \sum_{\{k:t_k^c < t_i^c, k \in C_i \text{ or } C_i^c\}} H(t_i^c|t_k^c; \alpha_{k,i}) \right] \quad (3)$$

As one may observe, the condition related to clusters in Eq. (1) and Eq. (2) are incorporated in the last two terms in Eq. (3), which are related to the potential influencers of node $i$. Comparing Eq. (3) to the original log-likelihood of NetRate, the additional cluster-related constraints can reduce the number of sub terms and parameters. This is especially critical to the efficiency of adopting gradient descent to estimate $\boldsymbol{A}$, since the last term in Eq. (3) has a "log of sum" form. Readers who are familiar with the gradient descent method may immediately recognize the gradient of this term is generally very expensive to compute during the parameter estimation process, which is essentially the bottleneck of the NetRate and other similar survival analysis based models.

Finally, similar to the original NetRate, we view each cascade as independent from each other. As a result, CENI adopts gradient descent to solve the following convex optimization problem, where the details of the computation can be found in related work (Gomez-Rodriguez, Balduzzi, and Schölkopf 2011; Gomez-Rodriguez, Leskovec, and Schölkopf 2013):

$$\begin{aligned}\underset{\boldsymbol{A}}{\text{minimize}} \quad & -\sum_c \ell_c(\boldsymbol{t}^c; \boldsymbol{A}) \\ \text{subject to} \quad & \alpha_{i,j} \geq 0, i, j = 1, 2, ..., N \end{aligned} \quad (4)$$

---

[1]Details of the computation can be found in (Gomez-Rodriguez, Balduzzi, and Schölkopf 2011)

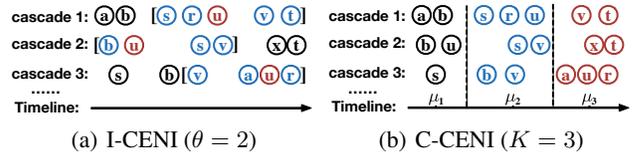

(a) I-CENI ($\theta = 2$)  (b) C-CENI ($K = 3$)

Figure 2: Clustering results of I-CENI and C-CENI. In (a), red nodes ($u$) are the centers of three time windows of size 2 ($\theta = 2$). The boundaries of time windows are marked by square brackets, and the blue nodes are considered in the same cluster with $u$ for each cascade. In (b), the boundaries of clusters are marked by dashed lines, and nodes in the same cluster share the same color. Different $\mu$s are the centroids of each cluster.

### Implementations of the Clustering Phase

In this article, we propose two implementations of the clustering phase, *Infection-centric CENI* and *Cascade-centric CENI*, which directly perform clustering on timelines of cascades. The third implementation of Phase 1, *Projection-based CENI*, is more sophisticated, which has an additional step of projecting nodes into a certain dimensional space before clustering.

**Infection-centric CENI (I-CENI)** Infection-centric CENI (I-CENI) imposes a time window to identify the cluster members for each node in each cascade. Assuming that all nodes are sorted in the ascending order of their infection time to cascade $c$, let $\tau(i, c)$ be the position of node $i$ in the sorted list. Moreover, we define $\tau(i, c) = \infty$, if $t_i^c = \infty$. For a specific cascade $c$, node $i \in C_j^c$, if and only if $|\tau(i, c) - \tau(j, c)| <= \theta$, where $\theta$ is a predefined threshold. One should notice that the "clusters" in I-CENI are not conventional clusters as in many clustering problems: they are essentially the "neighborhoods" of nodes in a cascade. Moreover, such neighborhoods naturally change with respect to cascades, since each cascade is an independent timeline.

For a clearer explanation, we use the example of node $u$ (marked by red color) in Fig. 2(a) to illustrate the clustering result of I-CENI. The example contains the results on the first three cascades, and the size of time window is 2 ($\theta = 2$). As one may observe, the nodes in $C_u$ (blue nodes) are different in three cascades. Due to the condition that $u$ can only be influenced by nodes that are infected earlier than itself, we only need to consider $\{s, r\}$ as the potential influencers of $u$ for cascade 1 when computing the likelihood in the inference phase. Similarly, for cascade 2 and 3, the potential influencers of $u$ are $\{b\}$ and $\{y, a\}$, respectively.

**Cascade-centric CENI (C-CENI)** Unlike I-CENI performs clustering from the perspective of each node, Cascade-centric CENI (C-CENI) adopts a different strategy. As a prerequisite, we make an assumption that cascades have a same temporal pattern propagating through users, which provides us the chance to utilize the information across different cascades to cluster nodes. Generally speaking, this temporal pattern refers to the probability that a node will be infected at a certain time point. On some datasets, this assumption may be violated, which consequently hurts the performance of C-CENI. However, it may work well on some other datasets, possibly because of the cascades in these datasets have more uniform propagation patterns. Later, in the section of introducing experiments, we will encounter both of these situations.

C-CENI adopts a K-means based strategy for clustering. We first view each $t_i^c$ as a data point on a one-dimensional space and assume that we have a predefined cluster number, $K$, which can be determined by cross validations. Initially, we randomly select $K$ centroids for each cluster. Next, at each iteration, we first assign each data point to the cluster of which centroid is the closest, then we recalculate the new centroid for each cluster by using the mean of data points in the cluster. We continue to do this until it converges. Similar to the "clusters" in I-CENI, a node's cluster members in C-CENI may also change with respect to cascades, since the infection time of this node may vary a lot for different cascades.

We use the clustering results of C-CENI (Fig. 2(b)) on the same example introduced previously to further explain how C-CENI works. In Fig. 2(b), we set $K$ to 3. The nodes in the same cluster are marked with the same color, and $\mu_1 \sim \mu_3$ are the centroids of 3 clusters. It is obvious to see that node $u$ belongs to the second cluster in the first cascade, while it belongs to the first and third cluster in the second and third cascade, respectively. Therefore, in the inference phase, C-CENI only deems $\{s, r\}$ as the potential influencers of $u$ in cascade 1. Similarly, the potential influencers of $u$ in cascade 2 and 3 are $\{b\}$ and $\{a\}$, respectively.

**Projection-based CENI (P-CENI)** Through experiments on real datasets (which will be introduced later), we found that although the extracted clusters from the clustering phases of I-CENI and C-CENI can help us shorten the running time of the inference phase, the F-measure of inferred edges on some datasets may be much worse than the results of state-of-the-art methods. Since I-CENI and C-CENI both cluster nodes directly on the timeline, we suspect the data on a one-dimensional space is insufficient to produce cluster structures preserving the performance of network inference algorithms. Inspired by this, we propose a more sophisticated implementation of the clustering phase, Projection-based CENI (P-CENI), which provides users an additional degree of freedom: the dimension of the nodes' representations. Higher dimension potentially provides the model more description power of the observed cascade data, which yields the derived clusters better help the network inference. Accordingly, a new question naturally arises: how many dimensions will be sufficient to carry the information contained in the cascade data? In general, finding the dimension to generate the optimal edge inference results (a.k.a. *critical dimension*) is challenging. In the following paragraphs, besides the implementation of the clustering phase of P-CENI, we also introduce a simple heuristic that helps us to find the critical dimension before actually running the inference phase. Although we can hardly prove the heuristic is guaranteed to find the critical dimension for all datasets, it is very efficient and practically works on the two real datasets we used to evaluate CENI models.

Unlike I-CENI and C-CENI directly perform clustering on the timelines, P-CENI first projects each node onto a space of certain dimension, say $D$. In other words, each node in the network will be endowed with unique $D$-dimensional coordinates. In fact, how to obtain the coordinates to explain the observed cascades is a relatively new problem and has several solutions very recently (Bourigault et al. 2014; Kurashima et al. 2014). The intuition of these solutions is that if the infection time of two nodes are relatively close in observed cascades, their positions in the projected space should be near enough to reflect this closeness. We argue that this problem is very different from the traditional network inference problem, since each edge in $\mathcal{E}$ represents the influence between two users, which is intrinsically asymmetric. For example, in a Twitter network, a celebrity may have a large impact on general public, yet not many people have a notable influence to him/her. Therefore, the symmetric distance measure on the projected space cannot fully uncover the actual influence between users. However, the obtained coordinates of each node can provide a general density estimation around it, which might be enough for the clustering phase of CENI. Other possible methods to project nodes in P-CENI is to adopt traditional dimension reduction algorithms, such as Isomap (Tenenbaum, De Silva, and Langford 2000). We feed the algorithm with the average difference of infection time between each pair of users, which is deemed as their distance on an unknown "high-dimensional" Euclidean space. The obtained results are the coordinates of all nodes in the reduced $D$ dimensional space.

In the proposed P-CENI, we adopt the method in (Bourigault et al. 2014) to map nodes because of its effectiveness and efficiency. If we denote the coordinates of node $i$ by $z_i$, and let $s_c$ be the first infected user of cascade $c$, the projection procedure of P-CENI is to find $\{z_i | i = 1...N\}$ that minimizes the following hinge loss function, where $\max(x, y)$ is a function returning the larger one between $x$ and $y$:

$$\sum_c \sum_{\{i | t_i^c < \infty\}} \sum_{\{j | t_j^c > t_i^c\}} \max\left(0, 1 - (\|z_{s_c} - z_i\|^2 - \|z_{s_c} - z_j\|^2)\right) \quad (5)$$

Similar to (Bourigault et al. 2014), we also adopt Monte Carlo approximation to get the solution that minimizes Eq. (5). For brevity, we will not introduce more details of this procedure, which can be found in (Bourigault et al. 2014). After obtaining each $z_i$, a K-means clustering algorithm is applied on the coordinates of nodes to obtain the cluster members for each node. Since each node $i$ corresponds to a unique $z_i$, its cluster members, $C_i$, does not change with respect to cascades in P-CENI, which is different from the cases in I-CENI and C-CENI.

In general, it is hard to find the critical dimension that can generate inferred edges with the best quality. This is because it might be impossible to theoretically validate whether the obtained clusters in the clustering phase will ensure a good result of the network inference, which is performed subsequently in the second phase. Moreover, since the network inference is much more time-consuming than the projection and clustering procedures, it will be ideal to find the critical dimension in the first phase. As a result, we have searched for different measures that may capture the critical dimension. Through experiments, we found that cluster-quality based measures, such as Davies-Bouldin Index (DBI), cannot determine the critical dimension. However, the value of objective function in Eq. (5) can be a suitable heuristic estimator in this task. In practice, this means that we can fix $D$ by simply using the dimension of which the computed node coordinates can generate the minimal hinge loss. A possible explanation to this observation is that metrics like DBI evaluate the compactness of the obtained clusters, which might be irrelevant to whether the clusters can ensure high-quality inferred edges. The objective function in Eq. (5), on the other hand, utilizes the infected timestamps of nodes, which are actually the same information sources as the network inference process. Therefore, the value of Eq. (5) may be more indicative when we try to find the critical dimension. Experimental results related to this will be covered in the next section.

## Experiments

### Datasets Description

We first describe the two datasets used in this article. Both of the two datasets are publicly available: the first one is retrieved from a MemeTracker (Leskovec, Backstrom, and Kleinberg 2009) dataset recording the memes and links of websites in the January of 2009[2], and the second one is extracted from a Sina Weibo dataset[3] (Zhang et al. 2013).

**MemeTracker**  MemeTracker is a project developed by Jure Leskovec et al. to map the daily news cycle by tracking the most frequently appeared quotes and phrases in numerous news and blog websites (Leskovec, Backstrom, and Kleinberg 2009). Networks retrieved from MemeTracker data are used in several previous related articles to evaluate their proposed methods (Gomez-Rodriguez, Balduzzi, and Schölkopf 2011; Gomez-Rodriguez, Leskovec, and Schölkopf 2013; Kurashima et al. 2014). In this dataset, each website is viewed as a node in the network. Websites can actively publish articles containing different memes, where each meme is deemed as a cascade. Moreover, each article may contain hyperlinks that refer to other related articles. Therefore, these hyperlinks can construct edges which we viewed as the ground truth. Based on the raw data of Jan., 2009, we first extract the 1,000 most active websites, a.k.a. websites that post most memes, and their published memes in Jan., 2009. The observation time window ($T$) is set to 1 week, and those memes that do not have enough time to be observed are simply discarded. In total, we obtain 2,762 different cascades and the infection time (time point when each website posts it) of each node. Moreover, through the hyperlinks posted with the memes, we obtain 4,323 directed edges connecting different nodes.

---
[2]http://www.memetracker.org/data.html
[3]http://arnetminer.org/Influencelocality

| #Datasets | *MemeTracker* | *Sina Weibo* |
|---|---|---|
| **#Nodes** | 1,000 | 1,000 |
| **#Edges** | 4,323 | 1,671 |
| **#Cascades** | 2,762 | 2,907 |
| **Maximal #infections in cascades** | 240 | 163 |
| **Average #infections in cascades** | 14.85 | 6.62 |

Table 1: Statistics of Datasets

**Sina Weibo**  Sina Weibo is a Twitter-like microblogging website that widely used in China. The second network we used to evaluate the CENI models is extracted from the raw data of a public Sina Weibo dataset published by (Zhang et al. 2013). Similar to the MemeTracker dataset, we first extract the 1,000 most active users (users who post most tweets) and their posted tweets in a specific month (Aug., 2012). These users are deemed as nodes in the network and they are connected through each other through "follow" relationship. If a user follows another user, s/he may see the tweets posted by the followee and further responds to it through retweet or repost action. As a result, the "follow" relationship among users forms the information network to empower the diffusion of tweets, which are viewed as cascades in this case. Based on the extracted 1,000 nodes, we obtain 1,671 directed edges, which are retrieved from the "follower-followee" network that published in the original dataset. Moreover, we set the observation time window ($T$) for each cascade, a.k.a. tweet, to 2 weeks. Those infections (include both direct and indirect responds) within 2 weeks since the initial post of the tweet are recorded, and any tweets do not have enough time to be observed are removed. At last, we get 2,907 different tweets and the infection time of each node in the network. Some basic statistics of the two datasets are summarized in Table 1. As we can see, both the maximal and the average numbers of infected nodes in cascades of MemeTracker are much larger than the Sina Weibo's. As one may expect, although the numbers of cascades in two datasets are similar, we will see that the running time of a network inference algorithm on these two datasets is very different.

### Baselines

In the experiments, we use two state-of-the-art network inference methods to serve as the baselines when evaluating the proposed I-CENI, C-CENI and P-CENI models.

**NetRate.**  Since the inference phase adopts a slightly modified NetRate model, it is reasonable to include the original NetRate (Gomez-Rodriguez, Balduzzi, and Schölkopf 2011) as a baseline. Similar to the inference phase of CENI, we also try out the transmission likelihood functions adopting exponential and Rayleigh distribution for NetRate model. The only exception is that the original NetRate does not have the conditions related to clusters in Eq. (1) and Eq. (2). In other words, NetRate considers all the infected nodes earlier than a specific node as its potential influencers, when computing the likelihood in Eq. (3).

**MMRate.**  Another baseline is the MMRate proposed by Wang et al. in (Wang et al. 2014). MMRate intends to utilize the hidden "aspects" of cascades to improve the effec-tiveness of the NetRate model. The term "aspect" is similar to the topics of diffused cascades, which are viewed as latent factors. MMRate assumes that cascades of different aspects have distinct propagation patterns and the parameter associated with an edge can change with respect to different aspects. Therefore, identifying the aspect of a cascade and distinguishing the different propagation patterns of aspects may potentially obtain more accurately inferred edges. MMRate adopts an Expectation-Maximization (EM) algorithm to estimate the aspect of each cascade and the edges in the network from the observed cascades. Although MMRate usually converges in just several iterations, we may expect it is more time-consuming comparing to the NetRate model because of the existence of the EM process.

|        | Running Time (seconds) | Precision | Recall | F-measure |
|--------|------------------------|-----------|--------|-----------|
| P-CENI | 8182                   | **0.058** | 0.140  | **0.082** |
| C-CENI | **7729**               | 0.038     | 0.128  | 0.059     |
| I-CENI | 8559                   | 0.043     | **0.233** | 0.073  |
| MMRate | 37328                  | 0.046     | 0.144  | 0.070     |
| NetRate| 16461                  | 0.045     | 0.168  | 0.071     |

Table 2: Results on the MemeTracker Dataset (exponential transmission likelihood)

|        | Running Time (seconds) | Precision | Recall | F-measure |
|--------|------------------------|-----------|--------|-----------|
| P-CENI | **9860**               | **0.037** | 0.219  | **0.064** |
| C-CENI | 10121                  | 0.018     | 0.162  | 0.032     |
| I-CENI | 10446                  | 0.021     | **0.322** | 0.039  |
| MMRate | 46078                  | 0.033     | 0.007  | 0.012     |
| NetRate| 18378                  | 0.027     | 0.249  | 0.049     |

Table 3: Results on the MemeTracker Dataset (Rayleigh transmission likelihood)

|        | Running Time (seconds) | Precision | Recall | F-measure |
|--------|------------------------|-----------|--------|-----------|
| P-CENI | 5877                   | 0.083     | 0.557  | 0.145     |
| C-CENI | **5415**               | 0.082     | 0.539  | 0.142     |
| I-CENI | 7358                   | 0.078     | **0.571** | 0.137  |
| MMRate | 21677                  | **0.085** | 0.565  | **0.148** |
| NetRate| 14404                  | 0.081     | 0.566  | 0.142     |

Table 4: Results on the Sina Weibo Dataset (exponential transmission likelihood)

## Experimental Results

**Overall Performance** We first introduce the overall performance of the proposed three CENI models and two baselines. Besides the running time, we also report the effectiveness of different methods in terms of the following three metrics, where $\hat{\mathcal{E}}$ denotes the set of inferred edges:

$$\begin{aligned}
\text{Precision} &= |\hat{\mathcal{E}} \cap \mathcal{E}|/|\hat{\mathcal{E}}| \\
\text{Recall} &= |\hat{\mathcal{E}} \cap \mathcal{E}|/|\mathcal{E}| \\
\text{F-measure} &= 2 \cdot (\text{Precision}^{-1} + \text{Recall}^{-1})^{-1}
\end{aligned}$$

Since F-measure considers both precision and recall of the inferred edges, the reported overall performance of different algorithms are under the condition that every model is tuned to generate the best F-measure. In order to conduct a fair comparison, all the models are implemented and solved by CVX solver, which is a Matlab-based convex problem solving tool developed by Michael Grant and Stephen Boyd (Grant and Boyd 2014; Grant and Boyd 2008). The results on the MemeTracker dataset when the transmission likelihood function takes the exponential form are listed in Table 2, while the results when the transmission likelihood function takes the Rayleigh form are listed in Table 3. Similarly, the results on the Sina Weibo dataset when transmission likelihood function takes the two forms are listed in Table 4 and Table 5, respectively.

In general, we found that it is harder to infer edges in the MemeTracker dataset than in the Sina Weibo dataset. A reasonable explanation to this is that the number of edges in MemeTracker is much larger than the number of edges in Sina Weibo, which makes that it is insufficient to use a similar number of cascades to infer the larger network. Additionally, we observe that every model takes a longer time to infer edges in the Sina Weibo dataset than the edges in the MemeTracker dataset. From Table 1 we can see that the numbers of nodes and cascades in the two datasets are the same or similar, yet the maximal and average length of cascades are very different. Therefore, it might be reasonable to conclude that the difference of the running time is caused by the length of cascades. As one may observe, CENI models are much more efficient than MMRate and NetRate: they only need around 20% ∼ 50% of the running time of MMRate and NetRate on the two datasets. Among the three CENI models, C-CENI is usually the fastest one except for the MemeTracker dataset when transmission likelihood is a Rayleigh distribution. I-CENI, on the other hand, is the most time-consuming CENI model, yet always generates the highest recall. Incorporating the latent topics of diffused cascades enable the inferred edges of MMRate has higher F-measure than NetRate on the Sina Weibo dataset. However, MMRate seems not stable, since it produces much worse inferred edges on the MemeTracker dataset when the transmission likelihood takes a Rayleigh form. Considering it takes much longer time than the original NetRate and CENI, we think the proposed CENI models may be more cost-effective to solve the network inference problem.

Although both I-CENI and C-CENI work generally well on the Sina Weibo dataset, they are relatively worse on the MemeTracker. P-CENI, on the other hand, is more stable and can produce edges of the highest F-measure in most cases. We set the number of iterations of the Monte Carlo simulation in the node projection step of P-CENI to 100,000, which produces robust results to be used to find the critical dimension and generate clusters. We iterates through 1 to 8 for the dimension used in the projection to find the critical dimension. Comparing to thousands or tens of thousands of seconds spent on the network inference, finding the critical dimension only cost us less than 5 minutes for each dataset. Later, we will show that the critical dimension for the Sina Weibo dataset is exactly 1, yet it is 2 for the MemeTracker dataset. This actually explains why the inferred edges of one-dimensional timeline based CENI models, a.k.a. I-CENI and C-CENI, may have better F-measure

|   | Running Time (seconds) | Precision | Recall | F-measure |
|---|---|---|---|---|
| P-CENI | 7724 | **0.075** | 0.466 | **0.130** |
| C-CENI | **7408** | 0.073 | 0.458 | 0.126 |
| I-CENI | 9247 | 0.060 | **0.541** | 0.108 |
| MMRate | 33965 | 0.073 | 0.476 | 0.128 |
| NetRate | 16219 | 0.066 | 0.527 | 0.117 |

Table 5: Results on the Sina Weibo Dataset (Rayleigh transmission likelihood)

on the Sina Weibo dataset than the MemeTracker dataset.

**Critical Dimension** Next, we show the experimental results to demonstrate the correlation between the value of the hinge loss function in Eq. (5) and the F-measure of the inferred edges. More particularly, with respect to P-CENI, we draw the curve of the hinge loss values obtained by projecting nodes onto the spaces of different dimensions and compare it with the curve of the F-measure of inferred edges. These curves under different numbers of clusters selected in the clustering phase of P-CENI are very similar, which essentially lead to the same conclusion we are about to make. Therefore, for brevity, we only demonstrate the curves under the number of clusters that can generate the highest F-measure of the inferred edges for each dataset. The results on the MemeTracker and Sina Weibo datasets are demonstrated in Fig. 3 and Fig. 4, respectively.

As shown in these figures, although the rise and fall patterns of the curve of hinge loss does not exactly corresponds to the patterns of the F-measure, the hinge loss does indeed successfully captures the dimension that can generate the best inferred edges in terms of the F-measure. To be more specific, the hinge loss of the projected nodes takes the minimal value when dimension is 2 on the MemeTrakcer dataset and 1 on the Sina Weibo dataset. Meanwhile, the F-measure of inferred edges (for both exponential and Rayleigh transmission likelihood) also takes the maximal value when dimension is 2 on the MemeTracker dataset and 1 on the Sina Weibo dataset. Since the critical dimension does not change due to the selected type of transmission likelihood, it is more likely to be an inherent character of the network itself, which is preserved in the observed records of cascades. To sum up, the results on the two tested datasets confirms that the value of Eq. (5) might be a good indicator to find the critical dimension in practice.

**Influence of the Change of $\theta$ and $K$**

In the last part of the experiment section, we illustrate how the F-measure of inferred edges by different CENI models changes with respect to their tunable parameters. For I-CENI model, the tunable parameter is the length of window ($\theta$), while the tunable parameters in C-CENI and P-CENI are the numbers of clusters ($K$). We only show the results of P-CENI when the dimensionality of projected space is exactly the critical dimension, since the critical dimension can be find before clustering. The results of different CENI models on the MemeTracker dataset are shown in Fig. 5 and the results on the Sina Weibo dataset are displayed in Fig. 6.

Generally speaking, the optimal value of $\theta$ or $K$ may vary

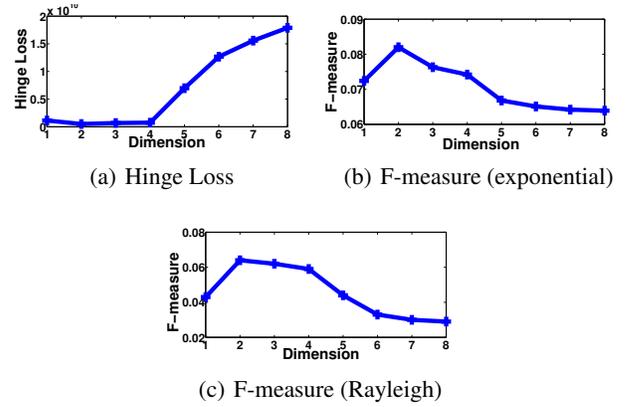

(a) Hinge Loss  (b) F-measure (exponential)

(c) F-measure (Rayleigh)

Figure 3: Comparison of the Hinge Loss of Node Projection and the F-measure of Inferred Edges under Different Transmission Likelihood Functions (MemeTracker Dataset)

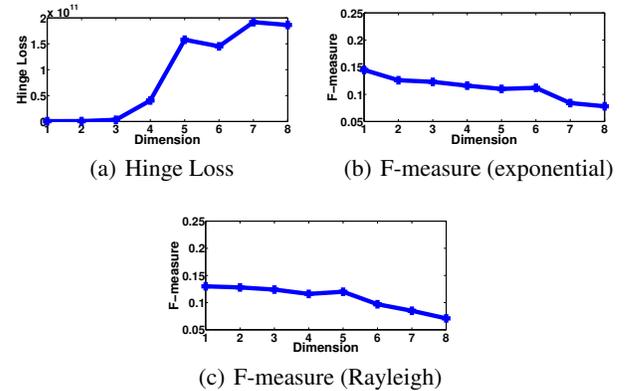

(a) Hinge Loss  (b) F-measure (exponential)

(c) F-measure (Rayleigh)

Figure 4: Comparison of the Hinge Loss of Node Projection and the F-measure of Inferred Edges under Different Transmission Likelihood Functions (Sina Weibo Dataset)

not only with the dataset, but also with the choice of transmission likelihood function. On the one hand, choosing the optimal $\theta$ for C-CENI prior to the inference phase is generally hard, since the length of cascades might vary a lot, and windows of an identical size may not always generate the best inferred edges on all datasets. On the other hand, for the choice of $K$, we find that traditional metrics evaluating the compactness of clusters like Davies-Bouldin Index cannot reveal the best $K$. This means it is also hard to find the optimal $K$ prior to the inference phases of C-CENI and P-CENI. Therefore, unlike the selection of the dimension used in the projection of P-CENI, we suggest users to utilize techniques such as cross validation to find a suitable $\theta$ or $K$ for a specific dataset.

As an interesting observation, we realize that the change of $K$ generally affects C-CENI more than P-CENI. This phenomena is especially notable on the Sina Weibo dataset when cascades are relatively shorter. By carefully comparing the clustering results of C-CENI and P-CENI, we find out that with the increase of the number of clusters, some P-CENI clusters may collapse into one cluster, which leads to the number of obtained clusters is smaller than the predefined $K$. This further causes the inferred edges under some large values of $K$ may have similar results to the ones under a smaller value of $K$. C-CENI, on the other hand, has a different story. In a general case, more nodes are usually infected shortly after the cascade starts than a long period later. It is hard to assign early infected nodes to reasonable clusters solely based on the timeline, since their timestamps of infection are too close. When the value of $K$ increases, we may get more clusters in the early phase of the cascade, which sacrifices too many inter-cluster edges, and thus causes the drop of the recall, as well as the F-measure.

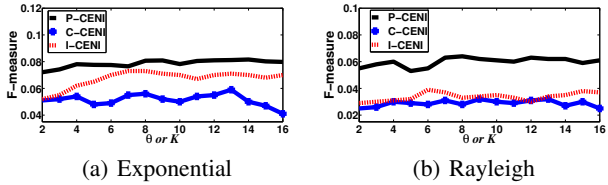

(a) Exponential    (b) Rayleigh

Figure 5: F-measure of Inferred Edges of CENI Models under Different Values of $\theta$ and $K$ (MemeTracker Dataset)

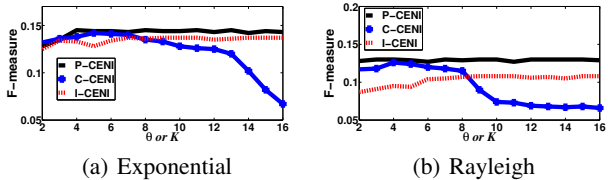

(a) Exponential    (b) Rayleigh

Figure 6: F-measure of Inferred Edges of CENI Models under Different Values of $\theta$ and $K$ (Sina Weibo Dataset)

## Related Work

The problem of inferring the hidden information network based on the observed cascades disseminated through it has been extensively studied recently. Based on several basic assumptions of *Independent Cascade* model (Kempe, Kleinberg, and Tardos 2003), (Gomez-Rodriguez, Leskovec, and Krause 2010) has developed NetInf model to solve the network inference problem. For NetInf, the edges in the network are viewed as identical, i.e. the connected source node has the same probability to infect the end node. NetInf iteratively adds the "most likely" existed edge to the set of inferred edges to uncover the unknown information network. The basic ideas of NetInf are also adopted in (Rodriguez and Schölkopf 2012), of which the proposed model admits an additional constraint that the hidden network is a forest. This condition relaxes the original network inference problem, which provides an important property, submodularity, to the objective function that required to be optimized. As a result, submodularity leads to an efficient greedy algorithm that guarantees a constant lower bound.

In reality, the infection probability associated with an edge naturally varies with the edge's source and end nodes. In order to capture such difference, NetRate (Gomez-Rodriguez, Balduzzi, and Schölkopf 2011) adopts three distinct parametric models that rooted in survival analysis to research the network inference problem. As an advantange against NetInf, NetRate is able to output not only the structure of the hidden network, but also the infection rate associated with each edge, which is governed by an associated model parameter. Based on the NetRate, various extended models are proposed to improve the effectiveness of it. Infopath (Gomez-Rodriguez, Leskovec, and Schölkopf 2013) applies a sampling step that biased to more recent cascades to generate the input to NetRate. As a result, the value of inferred parameters may change along with time, which better fits the dynamic evolution of the network. MoNet (Wang, Ermon, and Hopcroft 2012) incorporates the node dependent features into the transmission likelihood function of NetRate, which aims to increase the accuracy of inferred edges. Topic Cascade (Du et al. 2013), on the other hand, directly integrates the extracted topic distribution from the textual content of diffused cascades to infer the network, which is viewed as topic-sensitive. In other words, the infection rate associated with each edge changes with respect to different topics of the cascades. A more recent work, MMRate (Wang et al. 2014), is built on a similar idea as Topic Cascade with a fundamental difference that the topic of a cascade is viewed as a latent factor and needs to be estimated. However, same as the original NetRate, all these extended models do not scale well to the length of cascades: the number of infected nodes in cascades largely affects the running time of these models. However, in real life, we often encounter the situation that a contagious message is able to infect a large part of the network. This makes the existing methods become less ideal in these cases. The proposed CENI framework exactly tries to leverage the hidden cluster structures of nodes to improve the efficiency of the network inference process.

Another type of related works, such as (Bourigault et al. 2014; Kurashima et al. 2014), aim to find the coordinates of nodes in a specific Euclidean space, where the distances between nodes reflects their connection. To be more specific, if two nodes stay closer on the mapped space, they have a higher probability to influence each other, and their infection times in different cascades should be relatively closer as well. The model proposed by (Bourigault et al. 2014) attempts to minimize a hinge loss function by adopting Monte Carlo simulation to iteratively use stochastic gradient to adjust the coordinates of nodes. Meanwhile, (Kurashima et al. 2014) directly integrates the coordinates into the transmission likelihood of the NetRate and thus uses the same objective function as the optimization target. We deem such models actually do not fully solve the network inference problem: the distance between two nodes on the Euclidean space is symmetric, while the influence between them are intrinsically asymmetric. However, we think the obtained coordinates of a node can provide us indicative information of the density around it, which can be potentially useful for the clustering phase of the proposed P-CENI model. Therefore, we adopt the approach in (Bourigault et al. 2014) as a prerequisite step for the clustering of P-CENI due to its efficiency and effectiveness.

## Conclusion

In reality, we are often dealing with the case that we are aware of the time when each node responds to a message, while the connections among nodes causing the message's propagation are missing. Due to such connections are highly valuable to many real-life applications, such as viral marketing, the problem of unveiling the hidden network from the observed cascades recently has received extensive attention from researchers. Among proposed solutions, different survival analysis based approaches become increasingly popular and are widely used in many cases. However, we notice the efficiency of such models. To overcome this problem, this paper proposes a two-step framework, Clustering Embedded Network Inference (CENI), which incorporates the cluster structures of nodes to improve the efficiency of existing network inference algorithm. By adopting different clustering strategy, we propose three distinct implementations of CENI: Infection-centric CENI (I-CENI), Cascade-centric CENI (C-CENI) and Projection-based CENI (P-CENI). Furthermore, we point out the critical dimension problem: in order to ensure the obtained clusters can eventually produce high quality inferred edges, we may need to first estimate the dimensionality of the cascades and map the nodes to the space of that dimensionality. In order to ensure the efficiency of CENI framework, we adopt a hinge loss estimator to find the dimensionality before actually performing network inference process. Through substantial experiments based on two datasets, the results demonstrate that the proposed approaches only need around 20% $\sim$ 50% of the running time of the compared state-of-the-art approaches. Meanwhile, the inferred edges by the proposed approaches have similar or even slightly higher F-measure comparing to the edges inferred by the state-of-the-art algorithms.